\documentclass[man, 12pt, colorlinks=true, allcolors=cobalt]{apa7}
\definecolor{cobalt}{rgb}{0.0, 0.22, 0.61} 
\usepackage[english]{babel}
\usepackage{amsmath}
\usepackage{counttexruns}
\usepackage{graphicx}
\usepackage{float}
\usepackage{longtable} 
\usepackage{apacite}
\usepackage{natbib}
\usepackage{ragged2e} 
\usepackage{setspace} 
\usepackage{booktabs}     
\usepackage{array}        
\usepackage{caption}      
\usepackage{threeparttable} 
\usepackage{adjustbox}    
\usepackage{tabularx}     
\usepackage{pdflscape}    

\justifying

\title{Augmenting Coaching with GenAI: Insights into Use, Effectiveness, and Future Potential}
\shorttitle{AI-Augmented Coaching}

\author{Jennifer Haase}
\authorsaffiliations{Weizenbaum-Institute \& Humboldt Universität zu Berlin, Department for Computer Science}

\date{\today}

\newtheorem{researchquestion}{Research Question}
\makeatletter
\newcounter{subresearchquestion}
\let\savedc@researchquestion\c@researchquestion
\newcommand{\subresearchquestion}{
    \setcounter{subresearchquestion}{0}
    \stepcounter{researchquestion}
    \edef\saved@researchquestion{\theresearchquestion}
    \let\c@researchquestion\c@subresearchquestion
    \renewcommand{\theresearchquestion}{\saved@researchquestion\alph{researchquestion}}
}
\newcommand{\normresearchquestion}{
    \let\c@researchquestion\savedc@researchquestion
    \renewcommand\theresearchquestion{\arabic{researchquestion}}
}

\begin{document}
\vspace{-1em}  
\begin{center}
    \textbf{Pre-print}
\end{center}

\maketitle

\subsection*{Abstract}
The integration of generative AI (GenAI) tools, particularly large language models (LLMs), is transforming professional coaching workflows. This study explores how coaches use GenAI, the perceived benefits and limitations of these tools, and broader attitudes toward AI-assisted coaching. A survey of 205 coaching professionals reveals widespread adoption of GenAI for research, content creation, and administrative support, while its role in relational and interpretative coaching remains limited. Findings indicate that AI literacy and perceived AI impact strongly predict GenAI adoption, with positive attitudes fostering greater use. Ethical considerations, particularly transparency and data privacy, are a key concern, with frequent AI users demonstrating greater ethical awareness. Regression analyses show that while perceived effectiveness drives GenAI adoption, concerns about AI replacing human coaches do not significantly influence usage. Coaches express interest in future AI capabilities that enhance personalization, real-time feedback, and administrative automation while maintaining human oversight. The study highlights that GenAI functions best as an augmentation tool rather than a replacement, emphasizing the need for AI literacy training, ethical guidelines, and human-centered AI integration. These findings contribute to the ongoing discourse on human-AI collaboration, advocating for responsible and effective AI adoption in professional coaching.

\subsection*{Practice points}  
\begin{itemize}
    \setlength{\itemsep}{-2pt} 
    \setlength{\parskip}{0pt} 
    \item Coaches can use GenAI to enhance efficiency in research, content creation, and administrative tasks but should remain responsible for client relationships.  
    \item AI literacy is essential for ethical and effective AI use, helping coaches mitigate biases and integrate AI responsibly into their practice.  
    \item The future of coaching lies in Human-GenAI collaboration, where AI supports rather than replaces human expertise.  
\end{itemize}

\section{Introduction}

Coaching is a structured, goal-oriented, and collaborative process in which a coach facilitates the personal or professional development of an individual through targeted conversations, guidance, and strategic interventions. It is typically a non-directive approach that emphasizes self-discovery, accountability, and action, enabling individuals to enhance their skills, improve performance, and achieve specific objectives \citep{greif_coaching_2022}. Generative AI (GenAI) tools, particularly large language models (LLMs) such as ChatGPT and Gemini, have gained prominence for their ability to automate routine tasks \citep{huang_chatgpt_2023}, generate creative insights \citep{haase_artificial_2023}, and provide personalized feedback \citep{fuchs_exploring_2023}. Recent applications of GenAI tools have demonstrated their potential in coaching \citep{terblanche_artificial_2024}, where AI-driven chatbots effectively support self-reflection, stress management, and emotional regulation \citep{liu_is_2023, limpanopparat_user_2024}.

For coaching, these findings suggest diverse ways in which GenAI could transform the field. GenAI can (1) support the coach (e.g., by aiding education and efficient data analysis), (2) support the coaching process (before, during, and after sessions), and (3) emulate human coaches -- thus substituting them -- by "being" the coach through chatbots \citep{terblanche_artificial_2024, mai_chatbots_2023}. While a growing body of research highlights the effectiveness of 'AI Coaches' \citep{terblanche_coaching_2022, mai_best_2023}, particularly in improving accessibility and affordability \citep{kotte_digitalisierung_2023} and offering anonymity \citep{grasmann_coaching_2021}, significant reservations remain. These include the importance of human agency \citep{lenz_generative_2024}, the authenticity and credibility inherently tied to human coaches \citep{sheehan_paradox_2022}, and the current limitations of GenAI in mimicking empathy and fostering deep emotional connections \citep{webers_nicht_2024, mai_chatbots_2023}.

Rather than framing the discourse as a binary choice between GenAI as a coach vs. human coaches, it is worth exploring what a fruitful partnership between the two might look like. While existing research has emphasized the theoretical potential of GenAI and the capabilities of specific tools \citep{maiden_designing_2023, zachos_multi-technique_2022}, there remains considerable room to understand better how coaches already make use of these technologies. Some coaches reject the notion of GenAI support altogether \citep{diller_coach_2024}, while others embrace AI coaching assistants \citep{terblanche_influence_2024}. These contrasting perspectives raise critical questions: What can be learned from coaches who use GenAI tools? Which tools do they use? For which tasks do they employ these tools, and how do they evaluate their usefulness?

This paper addresses these questions by drawing on survey data from active coaching professionals. The study offers actionable insights into effectively integrating GenAI as a complementary tool. It argues that while GenAI can enhance efficiency and creativity, it must remain a supportive partner, preserving the inherently human aspects of coaching \citep{jarrahi_artificial_2018, webers_nicht_2024}.

\section{Foundations of Generative AI in Coaching}

\subsection{Basics of Generative AI and Large Language Models}
GenAI tools leverage vast datasets and sophisticated machine-learning algorithms to generate human-like text, automate repetitive tasks, and provide personalized outputs \citep{huang_chatgpt_2023}. Their versatility has enabled applications ranging from content creation to decision-making support, positioning them as valuable tools across industries. LLMs are built on a neural network architecture called the Transformer, which helps them process and generate text by analyzing how words in a sentence relate to each other. A key feature, the self-attention mechanism, allows the model to focus on the most important words, enabling it to produce coherent and contextually accurate responses \citep{raiaan_review_2024}. These models are pre-trained on extensive datasets, often comprising billions of words from diverse sources such as books, articles, and online content. This pre-training equips LLMs with a broad "understanding" of language, semantics, and contextual nuances \citep{kocon_chatgpt_2023}. Following the pre-training phase, LLMs undergo fine-tuning to adapt to specific tasks or domains. Fine-tuning involves exposing the model to smaller, domain-specific datasets to enhance its relevance and accuracy for particular applications, such as customer service, healthcare, or coaching \citep{li_chatdoctor_2023, mai_chatbots_2023}. This process ensures that LLMs can generalize their pre-trained knowledge while aligning their outputs with specific user needs and contexts.

The practical applications of LLMs span a wide range of domains due to their ability to generate coherent, context-aware text. In customer support, for instance, LLMs are employed to automate responses to common queries, improving efficiency and scalability \citep{davison_pickled_2023}. In healthcare, they assist in drafting clinical notes and supporting diagnostic processes for medical professionals \citep{li_chatdoctor_2023}. In education, LLMs are used for personalized tutoring, content generation, and language translation, enhancing learning experiences for students worldwide \citep{wu_ai_2024}. These capabilities also extend to creative domains, where LLMs aid in generating marketing copy, writing scripts, and brainstorming ideas \citep{doshi_generative_2024}.

\subsection{Generative AI and Coaching}

GenAI, particularly LLMs, has emerged as a transformative tool in coaching. These models can be applied to support tasks such as session preparation, creative resource generation, and client feedback analysis \citep{mai_chatbots_2023}. By synthesizing large amounts of information and generating contextually relevant outputs, LLMs enable coaches to streamline workflows and enhance the personalization of their services \citep{kotte_digitale_2024}. Additionally, conversational interfaces powered by LLMs facilitate real-time interactions, mimicking human dialogue to provide immediate and structured guidance. This versatility makes LLMs valuable tools for addressing the diverse and dynamic needs of coaching practices \citep{mai_chatbots_2023}.

The conceptual use of chatbots in a "therapy-like" context is not new. The world’s first chatbot, ELIZA, developed by J. Weizenbaum in 1966, simulated a Rogerian psychotherapist by responding to users' statements in a structured yet conversational manner \citep{weizenbaum_eliza_1966}. Modern GenAI tools build on this foundational idea, but with significantly more advanced capabilities. These tools' ability to provide structured, personalized feedback and scalable solutions has made coaching more accessible and cost-effective \citep{mai_chatbots_2023, kotte_digitalisierung_2023}. Furthermore, the anonymity and non-judgmental feedback offered by GenAI tools are especially valued by clients who prioritize privacy and convenience \citep{grasmann_coaching_2021}.

Despite its potential, using GenAI as a fully autonomous digital coach comes with significant downsides that limit its ability to replace human coaches. AI lacks emotional intelligence, genuine empathy, and the ability to foster deep client relationships—elements that are fundamental to effective coaching \citep{webers_nicht_2024, mai_chatbots_2023}. Coaching is not just about structured guidance but also about trust, intuition, and shared human experiences, which AI cannot replicate \citep{sheehan_paradox_2022}. Given these limitations, a purely AI-driven coaching model is unlikely to be a viable alternative to human coaches. Instead, integrating AI as a supportive tool --rather than a substitute-- may offer a more effective and ethically sound approach. This calls for a shift toward \textit{Human-GenAI collaboration}, where AI enhances but does not replace the human elements essential to coaching.

\section{Conceptual Framework of Human-AI Collaboration}

The concept of Human-AI Collaboration emphasizes the complementary relationship between human and AI capabilities, advocating for an integrated approach where both entities work together to enhance their respective strengths \citep{jarrahi_artificial_2018}. Within this framework, AI excels at tasks requiring data-driven precision, such as analyzing large datasets, recognizing patterns, and generating creative alternatives \citep{bewersdorff_myths_2023, haase_artificial_2023}. Humans, in contrast, contribute emotional intelligence, contextual understanding, and relational depth, which are indispensable for tasks requiring empathy and interpersonal engagement \citep{tankelevitch_metacognitive_2024, grigsby_artificial_2018}.

Building on the principles of human-centered AI, \cite{shneiderman_bridging_2020} highlights mainly two concepts essential for designing effective Human-AI Collaboration systems that are highly applicable to coaching. First, maintaining a balance between human oversight and AI automation is crucial to ensure that humans retain control over critical decisions and outcomes while AI contributes significantly to the coaching process. This balance preserves the integrity of coaching by aligning AI contributions with human judgment and relational depth. Second, AI tools in coaching should augment human capabilities, acting as powerful agents that expand creative and analytical potential. At this advanced level of collaboration, GenAI becomes an integrated partner, contributing unique ideas and insights that would not emerge through human effort alone. This approach underscores the transformative potential of human-AI collaboration in coaching by leveraging technology to enhance, rather than replace, human expertise.

Despite these opportunities, challenges persist in achieving seamless collaboration between humans and GenAI in coaching contexts. While GenAI tools can simulate empathy through advanced algorithms, they lack the emotional depth and authenticity necessary to build trust and foster meaningful client relationships \citep{webers_nicht_2024, mai_chatbots_2023}. Furthermore, the integration of GenAI into coaching workflows raises critical ethical and practical concerns. Issues such as biases embedded in training data, the opacity of AI decision-making processes, and the risk of over-reliance on technology underscore the importance of responsible implementation and oversight \citep{fuchs_exploring_2023}. Addressing these challenges is essential to fully realize the potential of human-AI collaboration in coaching, ensuring that the strengths of both humans and AI are utilized effectively while preserving the relational and ethical foundations of the practice.

%


\subsection{Ethical Dimensions of AI in Coaching}

As AI technologies become increasingly integrated into human workflows, they are evolving from passive tools to active social actors within organizational and relational contexts \citep{thiemann_maschine_2024}. In coaching, this shift is evident in the growing use of conversational interfaces that simulate dialogue and provide tailored recommendations, effectively mimicking human interactions \citep{mai_chatbots_2023}. While this evolution enhances the utility of AI in coaching, it also introduces significant ethical challenges. As AI systems take on more dynamic roles, questions regarding transparency, bias, and privacy become increasingly critical \citep{webers_nicht_2024}. Integrating AI into coaching workflows offers considerable advantages, such as improved efficiency and scalability, but it demands careful attention to these issues to ensure responsible use. For instance, although GenAI tools can deliver cost-effective and efficient solutions, their lack of transparency in decision-making processes can undermine trust between coaches and clients \citep{lenz_generative_2024}. Furthermore, biases embedded in AI algorithms may result in inappropriate or inequitable recommendations, underscoring the need for rigorous oversight and continuous refinement of these technologies \citep{webers_nicht_2024}.

Privacy concerns are particularly pronounced in coaching, where sensitive client information is routinely shared. Ensuring data confidentiality is not only a technical necessity but also a professional and ethical imperative \citep{lenz_generative_2024, webers_nicht_2024}. Coaches must adopt measures to protect client information when using GenAI tools, maintaining the trust that forms the foundation of effective coaching relationships. At the same time, they must balance the benefits of leveraging AI’s capabilities with the need to preserve the relational integrity of their practice. This highlights the importance of human oversight, where coaches remain accountable for the outcomes and ensure that AI tools align with ethical standards and the broader goals of coaching \citep{mai_best_2023}.

\subsection{Research Questions and Hypotheses}

While prior research highlights the theoretical promise of GenAI tools in automating tasks, generating creative insights, and providing personalized support \citep{jarrahi_artificial_2018, maiden_designing_2023}, empirical insights into real-world coaching applications remain scarce \citep{terblanche_artificial_2024, terblanche_influence_2024}. Additionally, concerns about ethical risks and professional boundaries require further investigation. Specifically, this study thus examines (1) how coaches use GenAI, (2) the perceived benefits and limitations of GenAI in coaching tasks, (3) how ethical and critical considerations shape GenAI adoption, and (4) what coaches wish AI could do better. 

\begin{researchquestion}
How do coaches currently use GenAI tools in their practice, and what factors influence adoption?
\end{researchquestion}

Technology adoption research emphasizes that digital literacy and attitudes toward the precise technology are key predictors of its use \citep{schepers_meta-analysis_2007, pan_technology_2020}. AI literacy, defined as the ability to understand, evaluate, and interact effectively with AI systems, has been shown to increase trust and reliance on AI in professional domains \citep{yang_how_2020}. Furthermore, attitudes toward AI significantly shape engagement, with positive attitudes increasing technology acceptance and negative attitudes fostering avoidance \citep{shao_understanding_2024}. Given this, the following hypotheses are proposed:

\begin{itemize}
    \setlength{\itemsep}{-2pt} 
    \setlength{\parskip}{0pt} 
    \item \textbf{H1:} Higher AI literacy is associated with greater GenAI use in coaching.
    \item \textbf{H2:} A more positive attitude toward AI is associated with greater GenAI use.
    \item \textbf{H3:} A more negative attitude toward AI is associated with lower GenAI use.
\end{itemize}

Research on GenAI tools like ChatGPT suggests that the technology shows great potential across most diverse domains, certainly with technology-affine domains like software design showing higher usage rates \citep{kocon_chatgpt_2023, rahmaniar_chatgpt_2023}.  Since GenAI can support coaching tasks in multiple specializations (e.g., business, career, and life coaching), it is expected that coaching type does not significantly influence usage patterns:

\begin{itemize}
    \setlength{\itemsep}{-2pt} 
    \setlength{\parskip}{0pt} 
    \item \textbf{H4:} Coaching specialization does not affect GenAI use.
\end{itemize}

\begin{researchquestion}
What are the perceived benefits and limitations of GenAI tools in supporting coaching tasks?
\end{researchquestion}

Perceived effectiveness has been established as a primary driver of sustained AI use \citep{schepers_meta-analysis_2007}. If GenAI is seen as useful, coaches are expected to integrate it more into their work \citep{ismatullaev_review_2024}. Additionally, GenAI users who use the tool often create a sense of the tool providing an important impact on the tasks at hand, which we know at least from educational settings \citep{chiu_motivating_2021}. Accordingly, the following hypotheses are formulated:

\begin{itemize}
    \setlength{\itemsep}{-2pt} 
    \setlength{\parskip}{0pt} 
    \item \textbf{H5:} Perceived effectiveness of GenAI is positively associated with GenAI use.
    \item \textbf{H6:} The role assigned to GenAI in coaching tasks is positively associated with GenAI use.
\end{itemize}


\begin{researchquestion}
How do ethical concerns and critical considerations shape coaches’ usage of GenAI?
\end{researchquestion}

Ethical AI adoption requires transparency, accountability, and human oversight \citep{shneiderman_bridging_2020}. Research in digital ethics suggests that frequent AI users tend to be more aware of ethical concerns, as they directly encounter issues such as bias, data privacy, and transparency \citep{ning_generative_2024}. Further, those using and knowing more about AI show less fear of being substituted by AI in their job, at least for radiologists \citep{huisman_international_2021}. Based on these findings, the following hypotheses are proposed:

\begin{itemize}
    \setlength{\itemsep}{-2pt} 
    \setlength{\parskip}{0pt} 
    \item \textbf{H7:} Greater use of GenAI is associated with higher attributed importance of ethical concerns.
    \item \textbf{H8:} Concerns about GenAI replacing coaching are negatively associated with AI and AI literacy.
\end{itemize}

Beyond assessing current usage patterns and perceptions, this study also explores how coaches envision the future role of GenAI in their profession. Understanding their expectations can provide valuable insights into emerging needs, potential challenges, and areas where AI-driven tools could offer meaningful support.  

\begin{researchquestion}
What do coaches wish to be supported with by GenAI?
\end{researchquestion}

\section{Methods}

This study uses an online survey to explore GenAI tool use in coaching practices, focusing on their applications, perceived effectiveness, and associated ethical considerations. The survey was designed to gather insights from active coaching professionals who integrate text-based GenAI tools, such as ChatGPT, Gemini, and Claude, into their work. The instrument combined closed-ended and open-ended questions to capture both quantitative and qualitative data. The aquestionnaire used and the raw-data can be found on OSF: \url{https://osf.io/fmazb/?view_only=d176140cf73c448680c564d1bf1e8c3b}

\subsection{Survey Structure and Measures}

The survey comprised several sections, each focussing on specific research objectives. Firstly, the professional background was recorded, followed by the exact use of the tools, the tasks they were used for, and how this collaboration was assessed. This was followed by a series of scales to record ethical behavior, AI literacy, and attitudes toward AI. In the end, the outlook for future developments was recorded. 

\subsubsection{Demographics and Professional Background.} Participants were asked about their professional roles (business, career, life, and health coach), years of coaching experience, and the GenAI tools they currently use. They also identified the tool they considered most important to their work, along with the frequency and duration of the tool use.

\subsubsection{AI Usage Patterns} The survey assessed how often participants used their primary AI tool for specific tasks of coaching work, see the following categories. The frequency was measured on a 5-point Likert scale ranging from "Never" to "Always".
\begin{itemize}
    \setlength{\itemsep}{-2pt} 
    \setlength{\parskip}{0pt} 
    \item \textbf{Creative and Resource-Building Tasks:} designing exercises, brainstorming solutions, creating personalized resources
    \item \textbf{Coaching Preparation and Support Tasks:} preparing session materials, tracking client progress, conducting research
    \item \textbf{Analytical and Reflective Tasks:} analyzing client data, identifying trends, summarizing takeaways
    \item \textbf{Relational and Empathy Tasks:} simulating supportive dialogue, providing empathetic feedback
    \item \textbf{Administrative and Operational Tasks:} scheduling, documenting sessions, organizing client files
    \item \textbf{Educational and Skill-Building Tasks:} learning new coaching techniques, training through AI simulations
\end{itemize}

\subsubsection{Perceived Effectiveness} Participants rated the effectiveness (Effect) of their primary GenAI tool in supporting tasks on a 5-point scale from "Ineffective" to "Transformative". Additionally, participants assessed the GenAI’s contribution to task outcomes (Role), categorized as "Minimal," "Supportive," "Collaborative," "Major," or "Complete". Participants were only presented with the tasks they previously claimed to use GenAI. 

\subsubsection{AI Literacy} AI literacy refers to the ability of coaches to understand the fundamental principles of AI, including the distinct roles of AI and humans in collaboration, the processes behind AI decision-making, and overall proficiency in utilizing AI tools effectively. It was assessed using a scale adapted from \cite{pinski_AI_2023}. Responses were recorded on a 7-point Likert scale ranging from "Strongly Disagree" to "Strongly Agree". All 3 items create a coherent score, with Cronbach’s~$\alpha = 0.820$.

\subsubsection{Attitudes Toward AI} General attitudes toward GenAI were assessed using items that explored both positive (e.g., excitement, perceived benefits; 12 items with Cronbach’s~$\alpha = 0.852$) and negative (e.g., concerns about misuse, discomfort; 8 items with Cronbach’s~$\alpha = 0.871$) perspectives. Responses were recorded on a 5-point Likert scale ranging from "Strongly Disagree" to "Strongly Agree" \citep{schepman_initial_2020}. 

\subsubsection{Ethical Considerations} Ethical considerations were measured through a series of items addressing key aspects of using GenAI tools in coaching practice. These included \textit{transparency with clients} (e.g., disclosing the use of AI tools and offering clients the option to decline AI-supported coaching), \textit{bias and fairness} (e.g., identifying and addressing biased outputs), \textit{privacy and confidentiality} (e.g., ensuring client data security), and \textit{responsibility and accountability} (e.g., monitoring AI outputs and aligning them with coaching goals). Additional items assessed the alignment of AI use with professional ethical standards, including adherence to the International Coaching Federation (ICF) code of ethics and concerns specific to GenAI, such as potential conflicts with professional judgment (cf. Table \ref{tab:ethics}). Responses were recorded on a 7-point Likert scale ranging from "Strongly Disagree" to "Strongly Agree". All 10 items create a coherent score, with Cronbach’s~$\alpha = 0.724$.

\subsubsection{Perceived Impact and Future Outlook of GenAI in Coaching} 
To assess participants’ perspectives on the broader implications of GenAI in coaching, several variables were measured. \textit{Perceived Current AI Impact} was assessed by asking participants to rate the extent to which GenAI tools currently influence their coaching practice, using a scale ranging from “No impact” to “Transformative impact.” \textit{Expected Future AI Use} measured participants’ anticipated reliance on GenAI in the future, with response options ranging from “Much less than now” to “Much more than now”. Additionally, two items captured concerns regarding the role of AI in coaching: \textit{AI Substitution Belief} evaluated the extent to which participants believed that GenAI tools could replace human coaching for individual tasks, rated on a scale from “Not at all” to “Completely”. \textit{AI Obsolescence Concern} assessed participants’ level of worry about GenAI making their role as a coach obsolete, ranging from “Not at all concerned” to “Extremely concerned”. To complement these quantitative measures, open-ended questions invited participants to elaborate on areas where they desired more effective AI support and to provide additional comments or suggestions regarding the role of GenAI in coaching.

\subsection{Procedure}

Participants were recruited through the \textit{Prolific} platform, targeting individuals who met two inclusion criteria: (1) actively working as a coach and (2) active use of GenAI tools in their coaching practice. Screening questions at the beginning of the survey ensured eligibility. 15 respondents who did not meet these criteria were excluded from the study. The survey began with an introduction explaining the purpose of the study and assuring participants of the confidentiality of their responses. Participants provided informed consent before proceeding. The survey took, on average 15 minutes to complete. Data collection was conducted in January 2025. Participants received 9 pounds per hour as monetary compensation. The Prolific platform provided descriptive data for the participants.

\subsection{Participant Description}

The survey included 205 participants, representing a diverse range of professional roles, demographic backgrounds, and engagement with GenAI tools in coaching. The majority of participants identified as health and fitness coaches (46.8\%), followed by business coaches (22.7\%), career coaches (19.7\%) and life coaches (10.8\%). The average participant age was 33.13 years ($SD = 11.69$, range = 19–76).

Gender distribution showed that 56.6\% identified as male and 42.4\% as female, with one participant preferring not to disclose gender. Most participants were based in the United States (55.1\%) or the United Kingdom (36.1\%), with the remaining 8.8\% residing in various European countries. Regarding employment status, 67.3\% worked full-time, 26.3\% worked part-time, and 0.5\% were unemployed but seeking work (cf. Table \ref{tab:descriptives}).

\begin{table}[h!]
\centering
\caption{Participant Descriptive Statistics}
\label{tab:descriptives}
\renewcommand{\arraystretch}{1.3} 
\begin{tabular}{lr}
\hline
\textbf{Characteristic} & \textbf{Frequency (\%)} \\ \hline
\textbf{Professional Role} &  \\
Business Coach & 46 (22.7) \\ 
Career Coach & 40 (19.7) \\ 
Life Coach & 22 (10.8) \\ 
Health and Fitness Coach & 95 (46.8) \\ \hline
\textbf{Gender} &  \\
Male & 116 (56.6) \\ 
Female & 87 (42.4) \\ 
Prefer not to say & 1 (0.5) \\ \hline
\textbf{Country of Residence} &  \\
United States & 113 (55.1) \\ 
United Kingdom & 74 (36.1) \\ 
Europe & 18 (8.8) \\ \hline
\textbf{Employment Status} &  \\
Full-time & 138 (67.3) \\ 
Part-time & 54 (26.3) \\ 
Unemployed (and job seeking) & 1 (0.5) \\ \hline
\textbf{Age} &  \\
Mean (SD) & 33.13 (11.69) \\ 
Range & 19–76 \\ \hline
\end{tabular}
\begin{tablenotes}
\vspace{6pt} 
\textit{Note:} Total (\textit{N}=205)
\end{tablenotes}
\end{table}

\section{Results}
\subsection{Descriptive Statistics and Preliminary Analyses}
Before testing the hypotheses, descriptive statistics were computed to summarize participant characteristics, GenAI usage patterns, and key study variables (see Tables~\ref{tab:descriptives} and \ref{tab:tool_use}). Pearson correlations were examined to explore initial relationships between GenAI use and evaluations, AI literacy, attitudes, and ethical considerations (see Table~\ref{tab:correlation_matrix}).

\begin{table}[h!]
\centering
\caption{Tool Use and Preferences}
\label{tab:tool_use}
\renewcommand{\arraystretch}{1.3} 
\begin{tabular}{lrr}
\hline
\textbf{Measure} & \textbf{Category} & \textbf{Frequency (\%)} \\ \hline
\textbf{Tools Used} & ChatGPT 3.5 & 137 (66.8) \\
 & ChatGPT 4.0 & 136 (66.3) \\
 & GitHub Copilot & 43 (21.0) \\
 & AlphaCode & 11 (5.4) \\
 & Gemini & 96 (46.8) \\
 & Claude & 34 (16.6) \\
 & BingChat & 32 (15.6) \\
 & Jasper AI & 20 (9.8) \\
 & Copy.AI & 24 (11.7) \\ 
 & Other GenAI tools & 5 (2.4) \\ \hline
\textbf{Most Frequently Used Tool} & ChatGPT 3.5 & 58 (28.3) \\
 & ChatGPT 4.0 & 134 (65.4) \\
 & Gemini & 8 (3.9) \\
 & Claude & 1 (0.5) \\
 & BingChat & 1 (0.5) \\
 & Jasper AI & 1 (0.5) \\
 & Other & 1 (0.5) \\ \hline
\textbf{Duration of Use} & More than 1 year & 102 (49.8) \\
 & 7–12 months & 55 (26.8) \\
 & 4–6 months & 28 (13.7) \\
 & 1–3 months & 16 (7.8) \\
 & Less than 1 month & 4 (2.0) \\ \hline
\textbf{Frequency of Use} & Multiple times per day & 74 (36.1) \\
 & Daily & 74 (36.1) \\
 & Several times a week & 35 (17.1) \\
 & Weekly & 14 (6.8) \\
 & Monthly or less & 8 (3.9) \\ \hline
\end{tabular}
\begin{tablenotes}
\vspace{6pt} 
\textit{Note:} Total (\textit{N}=205)
\end{tablenotes}
\end{table}

\begin{table}[h!]
    \centering
    \caption{Pearson Correlations Between Key Variables}
    \label{tab:correlation_matrix}
    \renewcommand{\arraystretch}{1.5} 
    \begin{adjustbox}{max width=\textwidth}
    \begin{tabular}{lcccccccccc}
        \hline
        & \rotatebox{90}{\textbf{Effect Meta-Score}} 
        & \rotatebox{90}{\textbf{Role Meta-Score}} 
        & \rotatebox{90}{\textbf{Ethics Score}} 
        & \rotatebox{90}{\textbf{AI Literacy Score}} 
        & \rotatebox{90}{\textbf{AI Positive Attitude}} 
        & \rotatebox{90}{\textbf{AI Negative Attitude}} 
        & \rotatebox{90}{\textbf{Current AI Impact}} 
        & \rotatebox{90}{\textbf{Future AI Use}} 
        & \rotatebox{90}{\textbf{AI as Substitute}} 
        & \rotatebox{90}{\textbf{AI Obsolescence Concern}} \\ 
        \hline
        \textbf{Use Meta-Score} 
        & .773\textsuperscript{**} & .721\textsuperscript{**} & .367\textsuperscript{**} & .567\textsuperscript{**} & .548\textsuperscript{**} & .037 & .589\textsuperscript{**} & .372\textsuperscript{**} & .267\textsuperscript{**} & .255\textsuperscript{**} \\
        \textbf{Effect Meta-Score}  
        &  & .755\textsuperscript{**} & .418\textsuperscript{**} & .575\textsuperscript{**} & .585\textsuperscript{**} & .052 & .533\textsuperscript{**} & .305\textsuperscript{**} & .238\textsuperscript{**} & .219\textsuperscript{**} \\
        \textbf{Role Meta-Score}  
        &  &  & .366\textsuperscript{**} & .467\textsuperscript{**} & .529\textsuperscript{**} & .233\textsuperscript{**} & .530\textsuperscript{**} & .316\textsuperscript{**} & .384\textsuperscript{**} & .338\textsuperscript{**} \\
        \textbf{Ethics Score}  
        &  &  &  & .379\textsuperscript{**} & .547\textsuperscript{**} & .087 & .231\textsuperscript{**} & .244\textsuperscript{**} & .087 & .144\textsuperscript{*} \\
        \textbf{AI Literacy Score}  
        &  &  &  &  & .552\textsuperscript{**} & -.146\textsuperscript{*} & .479\textsuperscript{**} & .396\textsuperscript{**} & .132 & .116 \\
        \textbf{AI Positive Attitude}  
        &  &  &  &  &  & -.040 & .465\textsuperscript{**} & .429\textsuperscript{**} & .127 & .101 \\
        \textbf{AI Negative Attitude}  
        &  &  &  &  &  &  & -.007 & -.070 & .379\textsuperscript{**} & .554\textsuperscript{**} \\
        \textbf{Current AI Impact}  
        &  &  &  &  &  &  &  & .580\textsuperscript{**} & .340\textsuperscript{**} & .235\textsuperscript{**} \\
        \textbf{Future AI Use}  
        &  &  &  &  &  &  &  &  & .217\textsuperscript{**} & .103 \\
        \textbf{AI as Substitute}  
        &  &  &  &  &  &  &  &  &  & .672\textsuperscript{**} \\
        \textbf{AI Obsolescence Concern}  
        &  &  &  &  &  &  &  &  &  &  \\
        \hline
    \end{tabular}
    \end{adjustbox}
    \begin{tablenotes}
    \vspace{6pt}
    \textsuperscript{**} Correlation is significant at the 0.01 level (2-tailed). \\
    \textsuperscript{*} Correlation is significant at the 0.05 level (2-tailed).
    \end{tablenotes}
\end{table}

\subsection{GenAI Use and Factors Influencing Adoption}

First, I analyzed how coaches use GenAI tools: they reported extensive use of GenAI tools, with \textit{ChatGPT 4.0} being the most frequently preferred tool across coaching practices (65.4\%). Regarding the duration of use, nearly half of the participants (49.8\%) had been using their preferred tool for more than one year, and most participants reported frequent interactions with their chosen tool, with 36.1\% using it daily and another 36.1\% using it multiple times per day (cf. Table~\ref{tab:tool_use}).

\subsubsection{Task-Specific GenAI Use, Perceived Effectiveness, and Assigned Role}

Participants reported their use of GenAI tools across different coaching-related tasks (Use), their perception of GenAI’s effectiveness in supporting these tasks (Effect), and the role they assigned to the AI in their coaching processes (Role) (cf. Table~\ref{tab:task_use_effect_role}).

\begin{table}[h!]
\centering
\renewcommand{\arraystretch}{1.2} 
\caption{GenAI Use, Effectiveness, and Assigned Role in Coaching Tasks}
\label{tab:task_use_effect_role}
\begin{threeparttable}
\begin{adjustbox}{max width=\textwidth} 
\begin{tabular}{lccc}
\hline
\textbf{Task Category} & \textbf{Use Mean (SD)} & \textbf{Effect Mean (SD)} & \textbf{Role Mean (SD)} \\ \hline
\multicolumn{4}{l}{\textbf{Creative \& Resource Tasks}} \\ 
Marketing content & 3.73 (0.99) & 3.83 (0.87) & 3.56 (1.05) \\
Design exercises & 3.84 (0.93) & 3.83 (0.90) & 3.49 (0.95) \\
Personalized resources & 3.75 (0.99) & 3.83 (0.89) & 3.41 (1.04) \\
Brainstorm solutions & 3.93 (1.00) & 3.80 (0.98) & 3.42 (1.07) \\ \hline
\multicolumn{4}{l}{\textbf{Preparation \& Support}} \\ 
Session materials & 3.90 (0.98) & 3.78 (0.87) & 3.55 (1.03) \\
Track progress & 3.44 (1.23) & 3.63 (1.01) & 3.33 (1.10) \\
Research topics & 4.01 (0.93) & 3.97 (0.81) & 3.76 (0.94) \\ \hline
\multicolumn{4}{l}{\textbf{Coaching Tasks}} \\ 
Set goals/plans & 3.72 (1.12) & 3.85 (0.83) & 3.47 (0.95) \\
Provide session materials & 3.73 (1.12) & 3.84 (0.93) & 3.48 (1.06) \\
Real-time problem-solving & 3.62 (1.16) & 3.79 (0.99) & 3.44 (1.02) \\ \hline
\multicolumn{4}{l}{\textbf{Analytical \& Reflective}} \\ 
Identify patterns & 3.60 (1.11) & 3.70 (0.99) & 3.49 (1.05) \\
Industry trends & 3.71 (1.12) & 3.90 (0.86) & 3.57 (0.96) \\
Summarize insights & 3.80 (0.99) & 3.76 (0.97) & 3.46 (1.01) \\ \hline
\multicolumn{4}{l}{\textbf{Admin \& Operations}} \\ 
Client communication & 3.69 (1.07) & 3.76 (0.94) & 3.40 (1.06) \\
Document summaries & 3.59 (1.12) & 3.72 (0.98) & 3.34 (1.07) \\
Automate scheduling & 3.59 (1.21) & 3.77 (0.87) & 3.53 (1.01) \\
Organize client files & 3.51 (1.30) & 3.78 (0.98) & 3.47 (1.05) \\ \hline
\multicolumn{4}{l}{\textbf{Relational \& Empathy}} \\ 
Refine listening & 3.61 (1.19) & 3.74 (1.00) & 3.40 (1.05) \\
Simulate dialogue & 3.50 (1.18) & 3.69 (0.98) & 3.45 (1.06) \\
Provide feedback & 3.67 (1.16) & 3.70 (0.93) & 3.39 (0.96) \\ \hline
\multicolumn{4}{l}{\textbf{Education \& Training}} \\ 
Learn new techniques & 3.90 (1.01) & 3.95 (0.91) & 3.66 (1.03) \\
Simulate training & 3.78 (1.14) & 4.03 (0.92) & 3.85 (0.97) \\ \hline
\textbf{Overall Meta-Scores} & 3.72 (0.80) & 3.76 (0.63) & 3.44 (0.74) \\ \hline
\end{tabular}
\end{adjustbox}
\begin{tablenotes}
\vspace{10pt} 
\parbox{0.99\textwidth}{ 
\textit{Note:} Overall \textit{N} = 205. 
Ratings were measured on 5-point Likert scales:  
\textit{Use}: 1 = Never, 2 = Rarely, 3 = Occasionally, 4 = Frequently, 5 = Always.  
\textit{Effect}: 1 = Ineffective, 2 = Marginally effective, 3 = Moderately effective, 4 = Highly effective, 5 = Indispensable.  
\textit{Role}: 1 = Minimal (AI had little impact), 2 = Supportive, 3 = Collaborative, 4 = Major, 5 = Complete (AI was the primary driver).  
} 
\end{tablenotes}
\end{threeparttable}
\end{table}

To assess whether the AI task usage items measured a single underlying construct, an exploratory factor analysis (EFA) was conducted using principal axis factoring. The Kaiser-Meyer-Olkin (KMO) measure of sampling adequacy indicated that the data was suitable for factor analysis (KMO = .954), and Bartlett’s test of sphericity was significant ($\chi^2(\text{df}) = 2944.343$, $p < .001$), confirming the presence of correlations among items. The first factor had an eigenvalue of 11.462, accounting for 52.1\% of the total variance, whereas the second factor had an eigenvalue of 1.344, indicating a steep drop in explanatory power. Given that the second factor’s eigenvalue was barely above 1 and contributed minimal variance, a single-factor solution was retained.

The Cronbach’s alpha for the AI task usage scale was 0.96, indicating excellent internal consistency. Given the strong factor loading structure and high reliability, I computed a composite AI Task Usage Score by averaging all task-related items. This composite score was used for subsequent analyses.

I repeated the analysis with \textit{effectiveness} and \textit{role}, with very similar patterns: For AI effectiveness, the KMO measure was .911, and Bartlett’s test was significant ($\chi^2(\text{df}) = 1608.687$, $p < .001$). The first factor had an eigenvalue of 8.828, explaining 40.1\% of the total variance, with a steep drop at the second factor (eigenvalue = 1.410). Reliability was high ($\alpha = .928$), justifying a single-factor solution. For the perceived role of AI, the KMO measure was .932, and Bartlett’s test was significant ($\chi^2(\text{df}) = 1862.148$, $p < .001$). The first factor (eigenvalue = 9.893) explained 44.96\% of the variance, with the second factor contributing minimally (eigenvalue = 1.367). Internal consistency was excellent ($\alpha = .941$), supporting a unidimensional structure. Across all three measures, results indicate that a single-factor structure is appropriate, justifying the use of composite scores for further analyses (see Table~\ref{tab:task_use_effect_role}).

\subsubsection{Predictors of GenAI Use}
A multiple regression analysis was conducted to examine the combined influence of AI literacy, AI attitudes, ethical considerations, perceived AI impact, expected future AI use, and concerns about AI substitution on GenAI use in coaching. The overall model was statistically significant (\(R^2 = 0.506\), \(\text{Adjusted } R^2 = 0.488\), \(F(8,196) = 25.343\), \(p < .001\)), explaining approximately 50.6\% of the variance in GenAI use (see Table~\ref{tab:regression_genai_full}).

Consistent with H1, \textit{Perceived Current AI Impact} emerged as the strongest predictor of GenAI use (\(B = 0.323\), \(p < .001\), \(\beta = 0.349\)), confirming that coaches with greater AI literacy are more likely to adopt GenAI tools in their practice. Supporting H2, \textit{AI Literacy} was also a significant predictor (\(B = 0.320\), \(p < .001\), \(\beta = 0.269\)), suggesting that higher AI literacy is associated with greater AI adoption. Similarly, \textit{Positive AI Attitudes} significantly predicted GenAI use (\(B = 0.308\), \(p = 0.002\), \(\beta = 0.216\)), reinforcing the role of AI acceptance in driving adoption.

In contrast, H3 was not supported, as \textit{Negative AI Attitudes} did not significantly predict GenAI use (\(B = 0.016\), \(p = 0.773\), \(\beta = 0.018\)). Similarly, \textit{Ethical Considerations} (\(B = 0.103\), \(p = 0.302\), \(\beta = 0.063\)), \textit{Expected Future AI Use} (\(B = -0.061\), \(p = 0.359\), \(\beta = -0.059\)), \textit{Belief in AI as a Substitute} (\(B = 0.016\), \(p = 0.730\), \(\beta = 0.024\)), and \textit{Concerns About AI Replacing Coaching} (\(B = 0.061\), \(p = 0.236\), \(\beta = 0.091\)) were not significant predictors of GenAI use.

Together, these findings provide strong empirical support for H1 and H2, emphasizing the critical role of AI literacy and positive attitudes in AI adoption among coaches. However, H3 was not supported, suggesting that negative attitudes toward AI do not significantly deter usage. Additionally, concerns about AI replacement, ethical considerations, and expectations for future AI use do not appear to drive or inhibit current GenAI adoption in coaching.

\begin{table}[h!]
    \centering
    \caption{Multiple Regression Analysis Predicting GenAI Use}
    \label{tab:regression_genai_full}
    \renewcommand{\arraystretch}{1.3} 
 \begin{tabular}{p{6cm} >{\centering\arraybackslash}p{1.5cm} >{\centering\arraybackslash}p{1.5cm} >{\centering\arraybackslash}p{1.5cm} >{\centering\arraybackslash}p{1.5cm} >{\centering\arraybackslash}p{1.5cm}} 
        \hline
        \textbf{Predictor} & \textbf{B} & \textbf{SE} & \textbf{Beta} & \textbf{t} & \textbf{p-value} \\
        \hline
        (Constant) & -0.434 & 0.391 & - & -1.111 & 0.268 \\
        AI Literacy Score & 0.320 & 0.077 & 0.269 & 4.130 & <.001 \\
        AI Positive Attitude Score & 0.308 & 0.100 & 0.216 & 3.068 & 0.002 \\
        AI Negative Attitude Score & 0.016 & 0.055 & 0.018 & 0.288 & 0.773 \\
        Ethics Score & 0.103 & 0.099 & 0.063 & 1.034 & 0.302 \\
        Perceived Current AI Impact & 0.323 & 0.064 & 0.349 & 5.063 & <.001 \\
        Expected Future AI Use & -0.061 & 0.066 & -0.059 & -0.920 & 0.359 \\
        AI Substitution Belief & 0.016 & 0.047 & 0.024 & 0.346 & 0.730 \\
        AI Obsolescence Concern & 0.061 & 0.051 & 0.091 & 1.189 & 0.236 \\
         \hline
        \textbf{Model Fit} & \multicolumn{5}{l}{\hspace{-3cm} $R^2 = 0.506$, $\text{Adjusted } R^2 = 0.488$, $F(8,196) = 25.343, p < .001$} \\
        \hline
    \end{tabular}
\end{table}


To assess whether coaching specialization influences GenAI adoption, a one-way ANOVA was conducted, comparing GenAI across tasks (meta-score) across all 4 types of coaching. H4 predicted that coaching specialization does not significantly affect GenAI use. The results revealed no significant differences --supporting the hypothesis-- between coaching types in GenAI Use ($F(3, 201) = 1.719, p = .134$) or Perceived Effectiveness ($F(3, 201) = 2.271, p = .081$), or for Role assigned to GenAI ($F(3, 201) = .371, p = .774$).


\subsection{Perceived Benefits and Limitations of GenAI Tools}

H5 predicted that the perceived effectiveness of GenAI would be positively associated with GenAI use. This hypothesis was supported (\(r = .773, p < .001\)), suggesting that the more effective coaches perceive GenAI to be, the more they use it. H6 predicted that the role assigned to GenAI in coaching tasks would be positively correlated with AI literacy and AI-positive attitudes. This hypothesis was supported (\(r = .467, p < .001\) for AI literacy; \(r = .529, p < .001\) for AI-positive attitudes), indicating that coaches with higher AI literacy and favorable AI attitudes tend to integrate GenAI more into their work (cf. Table~\ref{tab:task_use_effect_role}).


\subsubsection{Ethical and Critical Considerations}

The survey results indicate that ethical considerations play a crucial role in the integration of GenAI tools into coaching practices (see Table~\ref{tab:ethics}). 
Multiple linear regression was used to test whether GenAI use significantly predicted the attributed importance of ethical concerns in AI adoption. The overall regression was statistically significant (\(R^2 = 0.138\), \(F(1, 203) = 7.66\), \(p = .006\)), explaining 13.8\% of the variance in ethical considerations. GenAI use significantly predicted ethical considerations (\(\beta = 0.278\), \(p = .006\)), suggesting that frequent GenAI users are more aware of ethical issues --confirming H7. These findings are suggesting that frequent GenAI users are more attuned to ethical issues surrounding AI. This aligns with prior findings indicating that practitioners who regularly interact with AI systems are more likely to recognize and reflect on concerns related to transparency, bias, and accountability. Additional predictors, AI literacy, AI-positive attitudes and AI impact, were also examined but did not significantly contribute to ethical concern awareness in this model. This finding suggests that ethical engagement is more strongly linked to hands-on experience with AI rather than general AI knowledge or attitudes.

Multiple linear regression was used to test whether AI literacy and GenAI use significantly predicted concerns about AI replacing human coaching roles. H8 predicted that concerns about GenAI replacing coaching would be negatively associated with AI use and AI literacy. The overall regression was not statistically significant (\(R^2 = 0.035\), \(F(2, 202) = 2.21\), \(p = .111\)), suggesting that AI literacy and GenAI use do not explain significant variance in AI replacement concerns. GenAI use did not significantly predict AI replacement concerns (\(\beta = 0.064\), \(p = 0.211\)), and AI literacy also did not significantly predict AI replacement concerns (\(\beta = 0.116\), \(p = 0.098\)). Interestingly, while AI obsolescence concerns were strongly correlated with negative attitudes toward AI (\(r = .554, p < .001\)), they did not translate into reduced AI adoption. This suggests that while some coaches express conceptual concerns about AI replacing human roles, these fears do not influence their actual usage of GenAI tools.


\begin{table}[h!]
\centering
\renewcommand{\arraystretch}{1.3}
\caption{Ethical Considerations in AI-Supported Coaching}
\label{tab:ethics}
\begin{threeparttable}
\begin{adjustbox}{max width=\textwidth}
\begin{tabular}{lc}
\hline
\textbf{Ethical Consideration} & \textbf{Mean (SD)} \\ \hline
\textbf{Transparency} & \\
Disclose AI use to clients & 3.63 (1.13) \\
Clients can decline AI support & 3.61 (1.20) \\ \hline
\textbf{Privacy \& Accountability} & \\
Ensure client confidentiality & 4.37 (0.83) \\
Take responsibility for AI-supported outcomes & 4.24 (0.78) \\
Monitor AI output quality & 4.31 (0.77) \\ \hline
\textbf{Professional Alignment} & \\
AI aligns with ethical standards & 4.13 (0.78) \\
AI helps adhere to ICF ethics & 3.82 (0.97) \\ \hline
\textbf{Concerns About AI} & \\
Concerned AI conflicts with professional judgment & 3.34 (1.12) \\ \hline
\textbf{Human Oversight} & \\
Evaluate AI-generated content before sharing & 4.16 (0.80) \\ 
Confidence in identifying AI bias & 3.82 (1.08) \\ \hline
\end{tabular}
\end{adjustbox}
\begin{tablenotes}
\vspace{6pt} 
\textit{Note:} Total (\textit{N}=205), ratings measured on a 5-point Likert scale from 1 = Strongly Disagree to 5 = Strongly Agree.
\end{tablenotes}
\end{threeparttable}
\end{table}

\subsection{Desired Future Capabilities of GenAI in Coaching}

When asked about additional areas where they wished GenAI could support them, participants listed a variety of new features or further tasks GenAI should support with. To systematically analyze participants' open-ended responses regarding desired GenAI enhancements, a qualitative content analysis was conducted following an inductive coding approach \citep{mayring_qualitative_2015}. Responses were first reviewed for recurring themes, after which an initial codebook was developed. Through iterative refinement, codes were grouped into broader categories representing key areas where coaches seek AI support. This process ensured that emergent themes were grounded in the data while maintaining reliability in classification. I identified several key themes for AI enhancements in coaching (cf. Table~\ref{tab:ai_expectations}).

A major expectation was \textit{advanced client insights and behavioral analysis}, where AI could better analyze client emotions, recognize patterns in language and behavior, and predict potential challenges. Many coaches expressed interest in \textit{AI-driven personalized coaching plans }that integrate real-time data, cultural sensitivity, and multilingual capabilities. 
\textit{Administrative automation} remained a key demand, with requests for more efficient scheduling, session documentation, follow-up reminders, and AI-generated progress reports. Additionally, participants highlighted the need for AI to assist in content creation, including dynamic coaching exercises, interactive training modules, and automated resource generation. Another recurring theme was \textit{real-time coaching support}, where AI could provide live feedback, role-playing simulations, and adaptive coaching responses. Participants also emphasized the role of AI in business strategy and client management, envisioning AI helping with market analysis, financial planning, and advertising. Finally, several responses stressed the importance of \textit{ethical safeguards and data security}, calling for stronger protections for client confidentiality, bias detection, and adherence to professional coaching ethics. These findings suggest that while AI already supports many coaching workflows, there is significant potential for more sophisticated, context-aware, and ethical AI applications.  

\begin{table}[h!]
    \centering
    \caption{Coaches’ Desired Future GenAI Capabilities}
    \label{tab:ai_expectations}
    \renewcommand{\arraystretch}{1.3} 
    \begin{adjustbox}{max width=\textwidth} 
    \begin{tabular}{p{5cm}p{11cm}} 
        \hline
        \textbf{Key Expectation} & \textbf{GenAI should...} \\
        \hline
        \textbf{Client Insights} & ...analyze client progress, predict challenges, and provide emotional intelligence insights. \\
        \textbf{Personalization} & ...create hyper-personalized coaching plans that adapt dynamically and respect cultural differences. \\
        \textbf{Automation} & ...manage scheduling, session documentation, follow-ups, and coaching reports. \\
        \textbf{Content Creation} & ...generate interactive training modules, coaching exercises, and educational resources. \\
        \textbf{Real-Time Support} & ...offer live coaching assistance, role-playing simulations, and real-time feedback. \\
        \textbf{Business Strategy} & ...support financial planning, market research, and client management. \\
        \textbf{Ethical Safeguards} & ...ensure data security, detect biases, and align with coaching ethics. \\
        \hline
    \end{tabular}
    \end{adjustbox}
\end{table}

\subsection{Exploratory Analyses}

Beyond the primary hypotheses, additional analyses were conducted to explore emerging patterns related to ethical considerations, AI's impact as a moderator of AI attitudes, and future expectations for GenAI in coaching. These insights provide a deeper understanding of how coaches integrate AI into their practice, how their perceptions evolve with experience, and how AI literacy influences expectations for the future.

\subsubsection{Ethical Considerations and AI Use}

H7 predicted that greater GenAI use would be associated with higher attributed importance of ethical concerns. This hypothesis was supported (\( B = 0.278, p = .006 \)), suggesting that frequent GenAI users are more attuned to ethical considerations surrounding AI use. 
Examining correlation patterns (see Table~\ref{tab:correlation_matrix}), ethical considerations were positively associated with GenAI use (\( r = .367, p < .001 \)), AI literacy (\( r = .379, p < .001 \)), and positive attitudes toward AI (\( r = .547, p < .001 \)). Notably, ethical alignment also correlated with expectations of increased future AI use (\( r = .244, p < .001 \)), but showed no significant relationship with concerns about AI substituting human coaches (\( r = .087, p = .217 \)).  These findings challenge the assumption that ethical concerns primarily stem from automation fears. Instead, frequent AI users appear more focused on responsible AI integration, emphasizing transparency, bias detection, and professional accountability rather than resisting AI adoption. This aligns with previous research suggesting that ethical AI use depends on user awareness and engagement rather than outright skepticism \citep{ning_generative_2024}.

\subsubsection{Moderation Analysis: The Role of AI Impact}

To examine whether perceived AI impact moderates the relationship between AI attitudes and GenAI use, a series of moderation analyses were conducted using PROCESS Model 1 \citep{hayes_process_2012}. First, AI impact was tested as a moderator of positive AI attitudes predicting GenAI use. The interaction effect was significant (\( B = -0.2813, p < .001 \)), suggesting that the relationship between AI-positive attitudes and GenAI use \textit{weakens} as AI impact increases. Simple slopes analysis revealed that AI-positive attitudes strongly predicted GenAI use when AI impact was perceived as low, but the effect became non-significant at high levels of AI impact. 

Second, AI impact significantly moderated the effect of negative AI attitudes on GenAI use (\( B = -0.1355, p = .015 \)). Negative attitudes toward AI reduced GenAI use only when AI impact was perceived as low. When AI was seen as already having a high impact, negative attitudes did not significantly deter adoption. These findings suggest that AI adoption unfolds in stages. When AI is still an emerging concept, attitudes (both positive and negative) play a key role in shaping adoption. However, once AI is perceived as having a tangible impact, pragmatic factors such as effectiveness and usability become stronger predictors of usage than general attitudes. This aligns with research on technology adoption, which suggests that early adopters are often driven by attitudes, while later users rely more on observed utility \citep{escobar-rodriguez_acceptance_2014}. 

\subsubsection{Future AI Expectations: The Role of AI Literacy and Current GenAI Use}

A multiple linear regression analysis was conducted to examine whether \textit{AI literacy} and \textit{current GenAI use} predict anticipated \textit{future AI adoption}. The overall regression model was statistically significant (\( R^2 = 0.189 \), \( F(2, 202) = 23.467 \), \( p < .001 \)), explaining 18.9\% of the variance in expected future AI use. Both predictors significantly contributed to the model. \textit{AI literacy} was a strong predictor of future AI adoption (\( B = 0.312 \), \( p < .001 \), \( \beta = 0.272 \)), suggesting that coaches with greater AI literacy are more likely to anticipate increased AI integration in their future coaching practice. \textit{GenAI use} also significantly predicted future AI adoption (\( B = 0.210 \), \( p = .005 \), \( \beta = 0.218 \)), indicating that frequent GenAI users are more likely to expect an increase in AI usage. These findings suggest that both AI literacy and practical experience with GenAI contribute to optimistic expectations for future AI integration in coaching, reinforcing the role of knowledge and direct engagement in shaping AI adoption trajectories.

The exploratory analyses reveal several key insights: (1) frequent GenAI users report stronger ethical awareness, but their concerns focus on responsible AI use rather than automation fears; (2) AI attitudes matter for early adopters but lose predictive power when AI impact is high, suggesting a shift from attitudinal to pragmatic decision-making; and (3) AI literacy predicts optimism about future AI adoption, highlighting the role of knowledge in shaping long-term AI expectations, while current GenAI use does not necessarily translate into future optimism. These findings contribute to a more nuanced understanding of AI adoption in coaching, emphasizing the importance of knowledge, perceived impact, and ethical awareness in shaping AI-related decision-making.

\section{Discussion}

This study investigated how coaches integrate generative AI (GenAI) tools into their practice, the perceived benefits and limitations of these tools, and broader attitudes toward AI-assisted coaching. The findings indicate that GenAI is being actively used across all coaching-associated tasks, from different types of coaches, bei it business coaches, life coaches or health and fitness coaches. They use it primarily for research, automation, and content generation, while its role remains somewhat limited in deeply interpersonal and decision-making aspects of coaching. This suggests that AI adoption in coaching is task-specific, reinforcing previous research on AI-human collaboration \citep{jarrahi_artificial_2018} and highlighting the need to maintain a balance between automation and human oversight \citep{shneiderman_bridging_2020}.  

Additionally, this study provides empirical insights into the key drivers of AI adoption, ethical considerations, and the extent to which AI is viewed as a potential substitute for human coaching. A key finding is that perceived AI impact is a stronger predictor of GenAI adoption than general attitudes, suggesting that direct experience with AI plays a critical role in shaping professional engagement with the technology \citep{yang_how_2020}. This challenges common assumptions that attitudes alone drive AI adoption and instead underscores the importance of practical exposure and usability in determining whether professionals integrate AI into their work. These results offer a more nuanced perspective on GenAI’s role in coaching, countering automation fears and emphasizing how GenAI functions as an augmentation tool rather than a replacement for human expertise \citep{rony_i_2024}.

\subsection{The Role of Generative AI in Coaching}

As the study results show, GenAI has become an increasingly common tool in coaching workflows, with frequent use across a wide range of tasks. On average, reported usability was high, with particularly strong support for tasks related to knowledge retrieval, content generation, and brainstorming. Coaches reported even high usage for tasks \textit{during} a coaching session. This reinforces findings from prior research that emphasize GenAI’s strengths in structured problem-solving and ad-hoc knowledge-retrieval \citep{amershi_guidelines_2019, montag_propensity_2023}. In line with other professional domains, LLMs such as ChatGPT streamline information synthesis and structured output generation, enabling coaches to enhance efficiency and redirect their efforts toward higher-order cognitive and interpersonal aspects of coaching \citep{huang_chatgpt_2023}.  

Despite its increasing integration into coaching tasks, the study reveals some variation in GenAI adoption across different coaching tasks. While GenAI is widely used for structured, analytical, and administrative functions, its role remains more limited in tasks requiring emotional intelligence, deep relational engagement, and nuanced feedback. This demonstrates that GenAI, while effective in enhancing productivity, lacks human intuition, empathy, and contextual sensitivity, which are critical in relational professions \citep{webers_nicht_2024, sheehan_paradox_2022}. These results also extend the work of \citet{grasmann_coaching_2021}, who noted that while some clients valued AI-generated feedback for its consistency and neutrality, it did not replace the relational depth provided by human coaches. The findings suggest that while AI can function as a useful support tool, it does not substitute for all aspects of coaching work, like interpersonal dimensions that define effective coaching.

A key finding of this study is that concerns about AI replacing human coaching roles do not significantly predict lower GenAI adoption. While some coaches express automation fears, this does not translate into a reluctance to use GenAI in practice. This contradicts widespread assumptions that professionals resist AI due to job security concerns \citep{huisman_international_2021}. Instead, the results suggest that practitioners who actively use GenAI view it as a tool for augmentation rather than replacement. Even more so, when looking at the wishes coaches have for future GenAI tools, they often would want even more and intrusive tools able to mimic emotions and interpersonal nuances. These results are particularly relevant for discussions on GenAI’s role in professional coaching, as it suggests that concerns over AI-driven automation may be overstated in the coaching field. Rather than rejecting AI out of fear, coaches selectively integrate it where it adds value. However, those who assess the current impact of GenAI as high --measured as an overall perception rather than in relation to specific tasks-- also express higher concerns about future automation risks. This suggests that professionals who already perceive GenAI as transforming their work may expect potential long-term shifts in the profession. The fact that these concerns exist alongside high AI adoption further highlights the complexity of professional attitudes toward AI: coaches may simultaneously recognize AI’s value in augmenting their work while also feeling uncertain about the broader implications of automation \citep{rony_i_2024}. This perspective aligns with research on human-AI collaboration, which emphasizes that technology uptake is improved when the practical utility of the automation tool is focused on, rather than potential future substitution \citep{jarrahi_artificial_2018}. The correlation and regression analyses further support these conclusions. The strong associations between GenAI use, perceived effectiveness, and the role assigned to AI indicate that coaches integrate GenAI primarily when it demonstrably enhances efficiency --a common pattern for technology uptake overall \citep{pan_technology_2020}. Higher usage and a stronger attributed role align with high AI literacy, as those who use the tool frequently tend to feel more competent in leveraging its capabilities. Positive attitudes toward AI further reinforce this relationship, as does a high perceived impact of AI on coaching practices. However, interestingly, these factors do not necessarily translate into greater anticipated future usage of GenAI, suggesting that current adoption patterns are driven by immediate utility rather than long-term projections.

These findings contribute to a more nuanced understanding of AI adoption in coaching, emphasizing that direct experience with AI fosters engagement but does not necessarily resolve broader uncertainties about the future role of AI in professional practice. While GenAI supports efficiency and structured problem-solving, its integration into coaching remains contingent on its ability to complement, rather than replace, the deeply human aspects of the profession \citep{terblanche_influence_2024}. The findings suggest that the primary benefits of GenAI in coaching include increased efficiency, content automation, and structured knowledge retrieval. These results reinforce prior research emphasizing GenAI’s role in streamlining administrative processes, enhancing knowledge accessibility, and assisting with structured decision-making \citep{kotte_digitale_2024, doshi_generative_2024}. Particularly in tasks that require synthesis of information, brainstorming, and resource creation, AI was perceived as highly effective, aligning with research highlighting GenAI’s strengths in knowledge-intensive professions \citep{amershi_guidelines_2019, montag_propensity_2023}.  

Despite these advantages, several limitations temper the enthusiasm for AI integration. Transparency in AI use varies among coaches, with some explicitly disclosing GenAI-assisted processes while others adopt a more discretionary approach. These findings align with \citet{montag_propensity_2023}, who emphasized that trust in AI is closely linked to transparency and ethical alignment. While many coaches recognize the importance of responsible AI integration, there is still no standardized approach for AI disclosure in coaching, mirroring broader concerns in professional AI ethics \citep{shneiderman_bridging_2020}. Given that ethical awareness was positively associated with GenAI use, it appears that frequent users are more conscious of AI-related ethical considerations, yet best practices for implementation remain ambiguous. The mere fact that almost 94\% reported using a version of ChatGPT as their primary GenAI tool is problematic from an ethical perspective, as the tools are not set up for the secure handling of personal, sensitive data \citep{sebastian_privacy_2023}. This has practical implications for coaching organizations and AI governance. Rather than focusing on abstract ethical debates, institutions should prioritize AI literacy training that provides professionals with direct exposure to AI tools. This approach would enable responsible AI adoption through education rather than restriction.

Another key limitation lies in GenAI’s role in supporting human-centered coaching. While GenAI tools can facilitate structured dialogue and prompt reflection, their effectiveness in understanding complex interpersonal dynamics remains constrained \citep{webers_nicht_2024}. The relatively lower ratings for GenAI’s role in delivering personalized feedback suggest that practitioners remain cautious about delegating interpretative tasks to GenAI. While \citet{bender_dangers_2021} speculate that AI could eventually develop advanced relational capabilities, the present findings indicate that AI’s role in relational coaching remains limited, emphasizing the importance of preserving human oversight in client interactions.

\subsection{Implications for Coaching and AI Development}

The findings of this study suggest that while GenAI is already a valuable tool in coaching, its role remains as a collaborative tool supporting the human-centered coaching process. Future research should prioritize longitudinal studies that track how AI adoption evolves over time and how its perceived utility and ethical considerations shift as AI models become more advanced. Additionally, further investigation is needed into whether coaches shift their skill-set due to the intense collaboration with GenAI. As known from other fields fused with AI, human skills tend to deteriorate when less used as technology takes over critical aspects of work, as observed in aviation \citep{smith_human-automation_2020}.  

While AI already supports many coaching workflows, the qualitative findings of this study suggest that coaches seek more specialized AI tools that go beyond generic LLMs. Participants expressed interest in AI-driven behavioral insights, such as emotion recognition and language pattern analysis, that could provide a deeper understanding of client needs. Another frequently mentioned area for improvement involves dynamic, personalized coaching plans that can adapt in real time to individual coaching processes. Additionally, many coaches highlighted the need for improved administrative automation to further streamline workflows, reducing the time spent on documentation and session management. This suggests that the future of AI in coaching is not about replacing human coaches, but rather about refining AI tools to better support coaching tasks. Developers should focus on building coaching-specific AI solutions that integrate ethical safeguards, maintain data security, and align with professional coaching standards. From a practical standpoint, coaching organizations must take a proactive approach in preparing for the continued integration of AI by investing in structured training programs. These programs should not only equip coaches with the technical skills to integrate AI tools effectively but also encourage critical reflection on when and how AI should be employed in coaching workflows. The development of standardized AI-use guidelines is essential to ensure that AI adoption aligns with core coaching principles. These guidelines should emphasize transparency, ethical accountability, and responsible implementation, preventing over-reliance on AI in areas where human expertise remains indispensable \citep{shao_understanding_2024, rony_i_2024}. Given the study’s findings on AI literacy and adoption, structured AI training should incorporate ethical reflection exercises that challenge coaches to assess AI’s strengths and weaknesses and to define appropriate boundaries for its use.  

Moreover, AI developers and researchers should focus on expanding the portfolio of AI-supported coaching tools rather than simply refining GenAI models to mimic human interaction. The goal should not be to replicate human relational engagement but to provide a spectrum of AI-assisted options that cater to diverse coaching preferences and needs. Some clients may prefer the anonymity and accessibility of AI-driven support, while others might benefit from a human-AI teaming approach, similar to blended-learning models in education that balance automated guidance with personalized human interaction \citep{park_role_2024}. This approach acknowledges that different coaching approaches and client preferences require flexibility in AI integration rather than a one-size-fits-all model. Additionally, specialized GenAI tools tailored for professional coaching could offer significant advantages, particularly for coaches who seek to flexibly incorporate AI into their work without compromising data security and ethical standards \citep{shneiderman_bridging_2020}. Secure, customizable AI tools could allow coaches to maintain greater control over AI-generated content while ensuring compliance with confidentiality and professional guidelines. Future AI advancements should prioritize adaptive, context-aware features that enable meaningful human-AI collaboration while maintaining clear distinctions between AI-generated insights and human expertise. By broadening the available AI tools rather than narrowly focusing on imitating human coaching, AI development can support a more personalized and ethically sound integration of technology in coaching practices.  

Engaging coaching organizations, professional associations, and AI researchers in dialogue about best practices, ethical standards, and AI-assisted methodologies will help ensure that AI is integrated in ways that support professional growth rather than introduce unintended risks. Future research should also examine the client perspective more closely, exploring how AI-assisted coaching is perceived by those receiving coaching services and whether AI alters the perceived authenticity, trust, or efficacy of the coaching relationship.

\subsection{Limitations and Future Research}

While this study provides some insights into the integration of GenAI in coaching, several limitations should be acknowledged. First, the sample consists exclusively of coaches who already use GenAI, which certainly introduces a selection bias \citep{bethlehem_selection_2010}. This means that the study captures perspectives from AI-adopting professionals but does not account for those who reject or refrain from using AI in their practice. This comes with two implications: the results discussed here do not fully represent the broader coaching community, and the results might be overly optimistic about the usability of GenAI tools --because those perceiving a tool as useful tend to use it \citep{venkatesh_user_2003}. Future research should incorporate perspectives from non-users to explore barriers to adoption, skepticism, and alternative coaching philosophies that may resist AI integration.

Second, this study relies on self-report measures, which may introduce biases in how participants assess their own AI literacy, ethical awareness, and attitudes toward AI \citep{jahedi_advantages_2014}. It is possible that coaches overestimate their ability to critically evaluate AI-generated content or underreport ethical concerns due to social desirability bias \citep{tan_is_2021}. To address this limitation, future research could incorporate experimental or observational studies to assess AI literacy more objectively and evaluate whether AI training programs effectively enhance ethical decision-making in coaching.

Third, while this study provides a snapshot of current AI adoption in coaching, it does not capture the longitudinal dynamics of how coaching practices evolve as AI tools become more advanced. Given the rapid development of AI technologies, future research should examine how coaches adapt over time, whether AI-driven tools change coaching methodologies, and how coaching skills shift as tools become more embedded in professional workflows. Longitudinal studies could track the long-term effects of AI use on coaching effectiveness, professional identity, and ethical standards.

Finally, this study focuses primarily on the experiences of coaches, but future research should explore the perspectives of clients receiving AI-assisted coaching. Understanding how clients perceive AI-supported interventions, how it influences their trust in coaching, and whether AI alters the perceived authenticity of the coaching relationship are critical questions that remain largely unexamined \citep{terblanche_artificial_2024}. Addressing these gaps will provide a more comprehensive understanding of how AI is reshaping the coaching profession.

\subsection{Conclusion}

This study provides empirical insights into the integration of GenAI in coaching, highlighting its strengths across all sorts of coaching-associated tasks. The findings emphasize that AI should be positioned as a collaborative assistant rather than a substitute for human coaching. While GenAI enhances efficiency, knowledge retrieval, and content generation, it cannot replace human intuition, empathy, or ethical judgment. By leveraging AI’s automation and scalability while maintaining human oversight, coaches can optimize their practice without compromising the relational depth and ethical integrity essential to the profession. Future advancements should focus on refining AI’s contextual awareness and ethical safeguards, ensuring responsible and effective integration into professional coaching workflows.

\section{Acknowledgments}
Jennifer Haase's work was supported by the German Federal Ministry of Education and Research (BMBF), grant number 16DII133 (Weizenbaum-Institute).

\bibliographystyle{apacite}
\bibliography{references}

\end{document}